\documentclass{bioinfo}
\copyrightyear{2013}
\pubyear{2013}
\usepackage{cite} 
\usepackage{url}  
\usepackage{booktabs}
\usepackage{graphicx} 
\usepackage{tikz}
\usetikzlibrary{arrows}
\usepackage{pgfplots} 

\usepackage{ifpdf}
\usepackage{url}

\usepackage{xspace,amsmath,url}
\usepackage{color}
\usepackage{booktabs}

\newcommand{\exclude}[1]{}

\newif\iffigsinpdf
\figsinpdftrue

\newcommand{\q}{\phantom0}
\newcommand{\qq}{\phantom{00}}

\newcommand{\qc}{\phantom{0,}}

\newcounter{lnoc}

\newcommand{\lno}[1][0]{{\footnotesize\sffamily 
\ifnum#1=0
\stepcounter{lnoc} 
\ifnum\thelnoc<10
\phantom0%
\fi
\thelnoc
\else
\thelnoc.#1
\fi
}\>}

\begin{document}
\firstpage{1}
\title{Disk-based genome sequencing data compression}
\author[Szymon Grabowski, Sebastian Deorowicz, {\L}ukasz Roguski]{Szymon Grabowski\,$^{1}$, Sebastian Deorowicz\,$^{2}\footnote{To whom correspondence should be addressed}$, {\L}ukasz Roguski\,$^{3,4}$}
\address{$^{1}$Institute of Applied Computer Science, Lodz University of Technology, Al.\ Politechniki 11, 90-924 {\L}\'{o}d\'{z}, Poland \\
$^{2}$Institute of Informatics, Silesian University of Technology, Akademicka 16, 44-100 Gliwice, Poland \\
$^{3}$Polish-Japanese Institute of Information Technology, Koszykowa 86, 02--008 Warszawa, Poland\\
$^{4}$Centro Nacional de An\'{a}lisis Gen\'{o}mico (CNAG), Barcelona, Spain\\
}

\history{Received on XXXXX; revised on XXXXX; accepted on XXXXX}
\editor{Associate Editor: XXXXXXX}
\maketitle

\begin{abstract}
\section{Motivation:}
High-coverage sequencing data have significant, yet hard to exploit, redundancy.
Most FASTQ compressors cannot efficiently compress the DNA stream of large datasets, 
since the redundancy between overlapping reads cannot be easily captured in
the (relatively small) main memory.
More interesting solutions for this problem are disk-based~\citep{Y2011,CBJR2012},
where the better of these two, from Cox~{\it et al.}~(\citeyear{CBJR2012}),
is based on the Burrows--Wheeler transform (BWT) and
achieves 0.518\,bits per base
for a 134.0\,Gb human genome sequencing collection with almost 45-fold coverage.

\section{Results:}
We propose ORCOM (Overlapping Reads COmpression with Minimizers),
a compression algorithm dedicated to sequencing reads (DNA only).
Our method makes use of a conceptually simple and easily parallelizable idea of
minimizers, to obtain 0.317\,bits per base as the compression ratio, 
allowing to fit the 134.0\,Gb dataset into only 5.31\,GB of space.

\section{Availability:}
\url{http://sun.aei.polsl.pl/orcom} 
under a free license.

\section{Supplementary data:} 
available at Bioinformatics online.

\section{Contact:} \href{sebastian.deorowicz@polsl.pl}{sebastian.deorowicz@polsl.pl}

\end{abstract}


\section{Introduction}
It is well-known that the growth of the amount of 
genome 
sequencing data 
produced in the last years outpaces the famous Moore law 
predicting the developments in computer hardware~\citep{K2011,DG2013}.
Confronted 
with this deluge of data, we can only hope for better algorithms 
protecting us from drowning.
Speaking about big data management in general, there are two main 
algorithmic concerns:
faster processing of the data (at preserved other aspects, 
like mapping quality in de novo or referential assemblers) 
and more succinct data representations (for compressed storage or indexes).
In this paper, we focus on the latter concern.

Raw sequencing data are usually kept in FASTQ format, with two main streams:
the DNA symbols and their corresponding quality scores.
Older specialized FASTQ compressors were lossless, squeezing the DNA stream down 
to about 1.5--1.8\,bpb (bits per base) and the quality stream to 3--4\,bpb,
but more recently it was noticed 
that a reasonable solution for lossy compression of the qualities has 
negligible impact on further analyzes, e.g., referential mapping or variant calling performance~\citep{WAA2012,I2012,CMT2014}.
This scenario became thus immediately practical, with scores lossily 
compressed to about 1\,bpb~\citep{JRC2014} or less~\citep{YYB2014}.
Note also that Illumina software for their HiSeq 2500 equipment
contains an option to reduce the number of quality scores (even to a few),
since it was shown that the fraction of discrepant SNPs grows slowly with
diminishing number of quality scores in Illumina's CASAVA package
({\footnotesize\url{http://support.illumina.com/sequencing/sequencing_software/casava.ilmn}}).
It is easy to notice that now the DNA stream becomes the main compression 
challenge.
Even if higher-order modeling~\citep{BM2013} or LZ77-style compression~\citep{DG2011} 
can lead to some improvement in DNA stream compression, 
we are aware of only two much more promising approaches.
Both solutions are disk-based.
Yanovsky~(\citeyear{Y2011}) creates a {\em similarity graph} for the dataset,
defined as a weighted undirected graph with vertices corresponding
to the reads of the dataset. 
For any two reads $s_1$ and $s_2$ the edge weight between them 
is related to the ``profitability'' of storing $s_1$ and the edit script 
for transforming it into $s_2$ versus storing both reads explicitly.
For this graph its minimum spanning tree (MST) is found.
During the MST traversal, each node is encoded using the set of 
{\em maximum exact matches} (MEMs) between the node's read and the 
read of its parent in the MST.
As a backend compressor, the popular 7zip is used.
ReCoil compresses a dataset of 192M Illumina 36\,bp reads 
({\footnotesize\url{http://www.ncbi.nlm.nih.gov/sra/SRX001540}}), 
with coverage below 3-fold,
to 1.34\,bpb.
This is an interesting result, but ReCoil is hardly scalable; 
the test took about 14 hours on a machine with 1.6\,GHz Intel Celeron CPU 
and four hard disks.

More recently, Cox~{\it et al.}~(\citeyear{CBJR2012}) took a different approach, 
based on the Burrows--Wheeler transform (BWT).
Their result for the same dataset was 1.21\,bpb in less than 65 minutes, 
on a Xeon X5450 (Quad-core) 3\,GHz processor.
The achievement is however more spectacular if the dataset coverage 
grows.
For 44.5-fold coverage of real human genome sequence data
the compressed size improves to as little as 
0.518\,bpb\footnote{Actually in~\citep{CBJR2012} the authors report 
0.484\,bpb, but their dataset is seemingly no longer available 
and in our experiments we use a slightly different one.}
allowing to represent 
the 134.0\,Gbp of input data in 8.7\,GB of space.

In this paper we present a new specialized compressor for such data, 
ORCOM (Overlapping Reads COmpression with Minimizers), 
achieving compression ratios surpassing the best known solutions.
For the two mentioned datasets it obtains the compression ratios of 
1.005\,bpb and 
0.317\,bpb, respectively.
ORCOM is also fast, producing the archives in about 8 and 
77 minutes, respectively, 
using 8 threads on an AMD Opteron\texttrademark \ 6136 2.4\,GHz machine.

\begin{methods}
\section{Methods}

Let $s = s[0]s[1] \ldots s[n-1]$ be a string of length $n$ over an finite alphabet $\Sigma$ of size $\sigma$.
We use the following notation, assuming $0 \leq i \leq j < n$:
$s[i]$ denotes the $(i+1)$th symbol of $s$, $s[i \ldots j]$ the substring 
$s[i]s[i+1] \ldots s[j]$ (called a factor of $s$), 
and $s \circ t$ the concatenation of strings $s$ and $t$.

Our algorithm, ORCOM, 
follows the ancient paradigm of external algorithms:
distribute the data into disk bins and then process (i.e., compress) 
each bin separately.
Still, the major problem with this approach in reads compression
concerns the bin criterion: 
how to detect similar (overlapping) reads, in order to pass them into 
the same bin?
Our solution makes use of the idea of minimizers~\citep{RHHMY2004}, 
a late bloomer in bioinformatics, 
cf.~\citep{MFC2012,LKHYYS2013,CLJSM2014,WS2014,DKGDG2014}.
Minimizers are a simple yet ingenious notion.
The minimizer for a read $s$ of length $r$ is the lexicographically 
smallest of its all $(r-p+1)$ $p$-mers; usually it is assumed that $p \ll r$.
This smallest $p$-mer may be the identifier of the bin into 
which the read is then dispatched.
Two reads with a large overlap are likely to share the same minimizer.
In the next paragraphs we present the details of our solution.

Assume the alphabet size $\sigma = 5$ (ACGTN).
A reasonable value of $p$ is about 10, but sending each bin to a file on disk 
would require $5^{10} = 9.77$M files, which is way too much.
Reducing this number to $4^{10}+1$ (all minimizers containing at least 
one symbol \texttt{N}, which are rare, are mapped to a single bin, labeled N) 
is still not satisfactory.
We solved this problem in a radical way, using essentially
one\footnote{In fact, there is one extra file, with metadata, yet this one is 
of minor overall importance and we skip further description.}
temporary file for all the bins, using large output buffers. 
To fetch the bin data in a further stage, the file has to be opened to read
and the required reads extracted to memory from several locations of the file.

As DNA sequences can be read in two directions: forwards and backwards 
(with complements of each nucleotide), we also process each read 
twice, in its given and reverse-complemented form.
%
%
Additionally, we introduce a ``skip zone'', that is, do not look 
for minimizers in read suffixes of (default) length $z = 12$ symbols.
This allows for some improvement in read ordering in the next step.
The minimizers are thus sought over $2(r-z-p+1)$ resulting $p$-mers.
We call them {\em canonical minimizers}.

However, a problem with strictly defined minimizers is uneven bin distribution.
This not only increases the peak memory use, but also hampers parallel 
execution as the requirement for load balancing is harder to fulfill.
To mitigate these problems we forbid some canonical minimizers, namely 
those that contain any triple AAA, CCC, GGG or TTT or at least one N.
The allowed canonical minimizers are further called {\em signatures}, 
a term that we also used (with a slightly different definition) 
for the minimizers used in KMC~2, a $k$-mer counting algorithm~\citep{DKGDG2014}.

In the next step, when the disk bins are built, 
we reordered the reads in bins to move overlapping 
reads possibly close to each other.
From a few simple sort criteria tried out, the one that worked best 
was to sort the reads $s_i$, for all $i$, according to the lexicographical 
order of the string 
$s_i[j \ldots r-1] \circ s_i[0 \ldots j-1]$, 
where $j$ is the beginning position of the signature 
for the read $s_i$.
Such a reordering has a major positive impact on further compression.
The reason is that overlaping reads are with high probability
close to each other in the reordered array.
The size of the skip zone should be chosen carefully.
When too small, some signatures will be found close to the end of the read
and the first factor of the sorting criterion, $s_i[j \ldots r]$,
will be too short to have a good chance of placing the read among
those that overlap it in the genome.
On the other hand, with the zone being too long many truly overlapping reads
will be forbidden.


The last phase is the backend compression on bin-by-bin basis.
We devised a specialized processing method, which produces several 
(interleaving) streams of data, finally compressed with either 
a well-known context-based compressor PPMd~\citep{Sh2002}
or a variant of arithmetic coder~\citep{SM2010}%
\footnote{We use a popular and fast arithmetic coding variant by Schindler (\url{http://www.compressconsult.com/rangecoder/}), 
also known as a range coder.}.
How we process the bins in detail, including careful mismatch handling, 
is presented in the next paragraphs.

We maintain a buffer (sliding window) of $m$ previous reads, 
storing also the position of the signature in each read.
For each read, we 
seek 
the read from the buffer which maximizes the overlap.
The distance between a pair of considered reads 
depends on 
the number of elementary 
operations transforming one into another.
For example, if the pair of reads is: \\
\verb"  AACGTXXXXCGGCAT", \\
\verb"  CCTXXXXCGGCATCC", \\
where \verb"XXXX" denotes a signature, 
we match them after a (conceptual) alignment:\\
\verb"  AACGTXXXXCGGCAT", \\
\verb"    CCTXXXXCGGCATCC", \\
to find that they differ with 1 mismatch (\texttt{G} vs \texttt{C}) 
and 2 end symbols of the second read have to be inserted, 
hence the distance is $c_m \times 1 + c_i \times 2$, 
where $c_m$ and $c_i$ are the mismatch and the insert cost, respectively. 
The default values for the parameters are: $c_m = 2$ and $c_i = 1$, 
and they were chosen experimentally.
In our example, the final distance is thus $2 \times 1 + 1 \times 2 = 4$.
The read among the $m$ previous ones that minimizes such a distance, 
and is not greater than $\mathit{max\_dist}$, set by default to a half of read length, is considered 
a reference for the current read.

Next, the referential matching data are sent into a few streams.

\paragraph{Flags.} 
Values from 
$\{f_\text{copy}, f_\text{diss}, f_\text{ex}, f_\text{mis}, f_\text{oth}\}$, with the following meaning:
\begin{itemize}
\item $f_\text{copy}$ -- the current read is identical to the previous one,
\item $f_\text{diss}$ -- the read is not similar to any read from the buffer; 
more precisely, the similarity distance exceeds a specified threshold $\mathit{max\_dist}$,
\item $f_\text{ex}$ -- the read overlaps with some read from the buffer 
without mismatches (only its trailing symbols are to be encoded),
\item $f_\text{mis}$ -- the read overlaps with some read from the buffer 
with exactly one mismatch at the last position of the referenced read,
\item $f_\text{oth}$ -- the read overlaps with some read from the buffer, but not in a way 
corresponding to flags $f_\text{ex}$ or $f_\text{mis}$.
\end{itemize}

\paragraph{Lengths.} Read lengths are stored here (1 byte per length in the current 
implementation, but a simple byte code can be used to handle the general case).

\paragraph{The five streams: lettersN, lettersA, lettersC, 
lettersG, lettersT.} 
(These are used only if ``Flag'' is 
$f_\text{ex}$, $f_\text{mis}$ or $f_\text{oth}$.)

``LettersN'' stores ($i$) all mismatching symbols from the current read 
where at the corresponding position of the referenced read there is symbol \textit{N}, 
and ($ii$) all trailing symbols from the current read beyond the match 
(i.e., \textit{C} and \textit{C} in the example above).

``LettersX'', for $X \in \{A, C, G, T\}$, stores all mismatching symbols 
from the current read where at the corresponding position of the 
referenced read there is symbol $X$ (in our running example, the 
mismatching \textit{C} would be encoded in the stream ``lettersG''.
Note that the alphabet size for any ``LettersX'' stream is 4, that is, 
$\{A, C, G, T, N\} \setminus \{X\}$.

\paragraph{Prev.} 
(Used only if ``Flag'' is $f_\text{ex}$, $f_\text{mis}$ or $f_\text{oth}$.)
Stores the location (id) of the referenced read from the buffer.

\paragraph{Shift.} 
(Used only if ``Flag'' is $f_\text{ex}$, $f_\text{mis}$ or $f_\text{oth}$.)
Stores the offsets of the current reads against their referenced read.
The offset may be negative.
For our running example, the offset is $+2$.

\paragraph{Matches.} 
(Used only if ``Flag'' is $f_\text{oth}$.)
Stores information on mismatch positions.
For our running example, the matching area has 13 symbols, but 4 of them 
belong to the signature and can thus be omitted (as the signature's position 
in the current read is known from the corresponding value in the stream ``Shift'').
A form of RLE coding is used here. 
Namely, each run of matching positions (of length at least~1) 
is encoded with its length on one byte, and if the byte value is less than 255 
and there are symbols left yet, we know that there must be a mismatch 
at the next position, so it is skipped over.
``Unpredicted'' mismatches are encoded with 0.
For our example, we obtain the sequence: 1, 7 (match of length 1; omitted mismatch; match of length 11, which is 7 plus 4 for the covered signature's area).

\paragraph{HReads.}  
(Used only if ``Flag'' is $f_\text{diss}$.)
Here the ``hard'' reads (not similar enough to any read from the buffer) 
are dispatched.
They are stored almost verbatim: the only change in the representation 
is to 
replace the 
signature
with an extra symbol (.).
This helps a little for the compression ratio.

\paragraph{Rev.} 
Contains binary flags telling if each read is processed directly or first 
reverse-complemented.

\bigskip
Some of the streams 
are compressed with a strong general-purpose compressor, PPMd
(\url{http://compression.ru/ds/ppmdj1.rar}), 
using switches \textsf{-o4} \textsf{-m16m} (order-4 context model 
with memory use up to 16\,MB), 
others with our range coder (RC), also of order-4.
Namely, the streams ``Flags'' and ``Rev'' are compressed with order-4 RC, 
the stream ``LettersX'' with order-4 RC, where the context is formed of the 
four previous symbols,
and all the other streams are compressed with PPMd.

The description presented above is somewhat simplified.
We took some effort to achieve high processing performance.
In particular, the input data (read from FASTQ files, possibly gzipped) are
processed in 256\,MB blocks (block size configurable as an input parameter)
and added to a queue.
Several worker threads find signatures in them, perform the necessary
processing and add to an output queue, whose data are subsequently
written to the temporary file.
Also further bin processing is parallelized, to maximize the performance.

\section{Results}

We tested our algorithm versus several competitors 
on real and simulated datasets, detailed in Table~\ref{tab:datasets}.
The test machine was a server equipped with four 8-core 
AMD Opteron\texttrademark \ 6136 2.4\,GHz CPUs, 128\,GB of RAM 
and a RAID-5 disk matrix containing 6 HDDs.
We use decimal multiples for the units, i.e., ``M'' (mega) is $10^6$,  
``G'' (giga) is $10^9$, etc., for the file sizes and memory uses 
reported in this section.

\subsection{Real datasets}
\begin{table}
\processtable{Characteristics of the datasets used in the experiments.\label{tab:datasets}}%
{
\renewcommand{\tabcolsep}{0.32em}
\begin{tabular}{lccccc}\toprule
Dataset 						&  Genome	& No.\	& No.\	& Avg. read	& Accession \\
(Organism)						&  length	& Gbases	& Mreads	& length		& no.			\\
\midrule
\emph{G.\ gallus}			&	1,040		& \q34.7	& \qc347	& 100			& SRX043656	\\
\emph{M.\ balbisiana}	&	\qc472	& \q56.9	& \qc569	& 101			& SRX339427	\\
\emph{H. sapiens} 1		&	3,093		& \qq6.9	& \qc192 & \q36		& SRX001540	\\
\emph{H. sapiens}	2 		& 	3,093		& 135.3	& 1,340	& 101			& ERA015743	\\
\emph{H. sapiens}	2-trim& 	3,093		& 134.0	& 1,340	& 100			& ERA015743	\\
\midrule
\emph{H.\ sapiens} 3		&	3,093		& 125.0	&	1,250	& 100			& ---\\
\emph{H.\ sapiens} 4		&	3,093		& 125.0	&	1,250	& 100 		& ---\\
\bottomrule
\end{tabular}
}{
Approximate genome lengths are in Mbases according to \url{http://www.ncbi.nlm.nih.gov/genome/}.
\emph{H.\ sapiens} 3 and 4 data sets are artificial reads produced by sampling reference human genome.
In \emph{H.~sapiens}~3 the sampled reads are exact and in \emph{H.~sapiens}~4 bases are modified with probability 1\%.}
\end{table}

\begin{table*}
\processtable{Compression ratios for various data sets\label{tab:results}}%
{
\begin{tabular}{@{\extracolsep{1.1em}}lcccccccc}
\toprule
Dataset							& DSRC~2	& Quip	& FQZComp& Scalce	& SRcomp	& ReCoil & BWT-SAP& ORCOM \\
\midrule
\emph{G.\ gallus} 			& 1.820	& 1.715	& 1.419	& 0.824	& 1.581	& ---		& 0.630	& \bf 0.433 \\
\emph{M.\ balbisiana}		& 1.838 	& 1.196	& 0.754 	& 0.342	& 0.522	& ---		& 0.208	& \bf 0.110 \\
\emph{H.\ sapiens} 1			& 1.857	& 1.773	& 1.681	& 1.263	& 1.210	& 1.34	& 1.246	& \bf 1.005\\
\emph{H.\ sapiens} 2			& 1.821	& 1.665	& 1.460	& 1.117	& NS		& ---		& NS	& \bf 0.327 \\
\emph{H.\ sapiens} 2-trim	& 1.839	& 1.682	& 1.474	& 0.781	& failed	& ---		& 0.518	& \bf 0.317 \\
\midrule
\emph{H.\ sapiens} 3 		&  1.832	& 1.710	& 1.487	& 0.720	& failed	& ---		& 	0.410	&\bf 0.174	\\
\emph{H.\ sapiens} 4 		&  1.902	& 1.754	& 1.568	& 1.022	& failed	& ---		& 	0.810	&\bf 0.562 \\
\bottomrule
\end{tabular}}%
{Compressed ratios, in bits per base. 
The results of our approach are presented in the rightmost column.
The best results are in bold.
`NS' means that the compressor was not examined as it does not support 
variable-length reads in a dataset.
ReCoil was not examined in our experiments due to very long running times (the only result comes from~\citep{CBJR2012} paper).
}
\end{table*}

\begin{table*}
\processtable{Compression times and memory usage of compressors\label{tab:results2}}%
{\renewcommand{\tabcolsep}{0.3em}
\begin{tabular}{@{\extracolsep{0.4em}}lccccccccccccccccccccc}
\toprule
Dataset					&& 
\multicolumn{2}{c}{DSRC~2}	&& 
\multicolumn{2}{c}{Quip}	&&
\multicolumn{2}{c}{FQZComp}&&
\multicolumn{2}{c}{Scalce}	&&
\multicolumn{2}{c}{SRcomp}	&&
\multicolumn{2}{c}{BWT-SAP}&& 
\multicolumn{2}{c}{ORCOM} \\
\cline{3-4}\cline{6-7}\cline{9-10}\cline{12-13}\cline{15-16}\cline{18-19}\cline{21-22}
&& time & RAM	&& time & RAM	&& time & RAM	&& time & RAM	&& time & RAM	&& time & RAM	&& time & RAM	\\
\midrule
\emph{G.\ gallus} 			&& 1.3	& 5.2	&&\q9.7& 0.8	&& 12.3& 4.2	&& \q4.6	&	5.5	&& 4.2	& 34.3&&\q62.3& \bf 0.003	&& \bf 1.1	&\q9.6 \\
\emph{M.\ balbisiana}		&& 2.2	& 5.4	&&	14.4& 0.8	&& 18.0& 4.2	&& \q9.7	& 5.4		&& 3.6	& 55.5&&\q99.1& \bf 0.003	&& \bf 2.1	&\q5.2	\\
\emph{H.\ sapiens} 1			&& \bf 0.3	& 6.1	&&\q1.6& 0.8	&&\q2.3& 4.2	&& \q1.2	& 5.5	&& \bf 0.3	&\q6.9&&\qq6.0& \bf 0.003	&& 0.5	&\q9.7	\\
\emph{H.\ sapiens} 2			&& 9.7	& 2.8&& 39.0& \bf 0.8	&& 47.4	&	4.3	&& 18.5& 5.5	&& \multicolumn{2}{c}{NS}		&& \multicolumn{2}{c}{NS}	&& \bf 5.6	&	13.8	\\
\emph{H.\ sapiens} 2-trim	&& 9.7	& 2.8	&& 39.0& 0.8	&& 45.9& 4.2	&& 18.5& 5.5	&& \multicolumn{2}{c}{failed}	&&267.8& \bf 0.003	&& \bf 4.6	& 12.5\\
\midrule
\emph{H.\ sapiens} 3 		&& \bf2.2	& 5.2	&&	35.3& 0.8	&& 45.0& 4.2	&& 12.8& 5.4	&& \multicolumn{2}{c}{failed} && 246.1	& \bf0.005	&& 2.8	& 11.7	\\
\emph{H.\ sapiens} 4 		&& \bf2.8	& 5.6 &&	42.7 &0.8	&& 48.7& 4.2	&&	16.0& 5.5	&& \multicolumn{2}{c}{failed}	&&	278.0	&\bf0.006		&&	5.0	&	13.7	\\
\bottomrule
\end{tabular}}%
{
Times are in thousands of seconds.
RAM consumptions are in GBs.
The best results are in bold.
`NS' means that the compressor was not examined as it does not support 
variable-length reads in a dataset.
}
\end{table*}

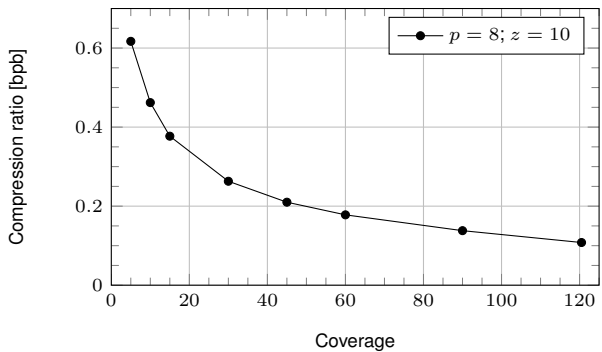
\begin{figure}[t]
\iffigsinpdf
\includegraphics{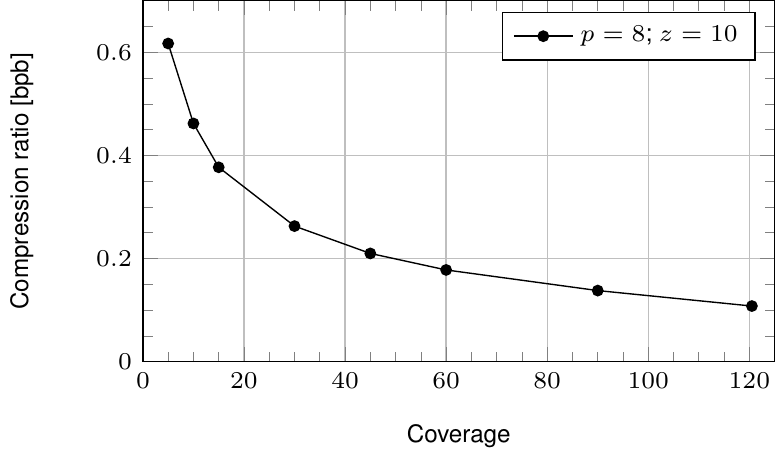}
\else
\centering
\pgfplotsset{width=8.0cm, height=5.25cm}
{\sffamily
\begin{tikzpicture}
\begin{axis}[
	xlabel={Coverage}, 
	ylabel={Compression ratio [bpb]},
	minor y tick num=3,
	minor x tick num=3,
	xmin=0,
	ymin=0,
	ymax=0.7,
	xmax=125,
	grid=major,
	mark size=1.5,
	font=\scriptsize,
	legend entries={$p = 8$; $z  = 10$},
	legend columns=1,
	legend pos={north east},
	legend style={font=\scriptsize, mark size=1.5},
]
\addplot[black, mark=*] table[x=Coverage, y=Ratio] {dat/coverage.dat};
\end{axis}
\end{tikzpicture}
}\fi
\caption{Compression ratio for various coverage factors for \emph{M.\ balbisiana} dataset}
\label{fig:res:time_k}
\end{figure}

\begin{figure}[t]
\iffigsinpdf
\includegraphics{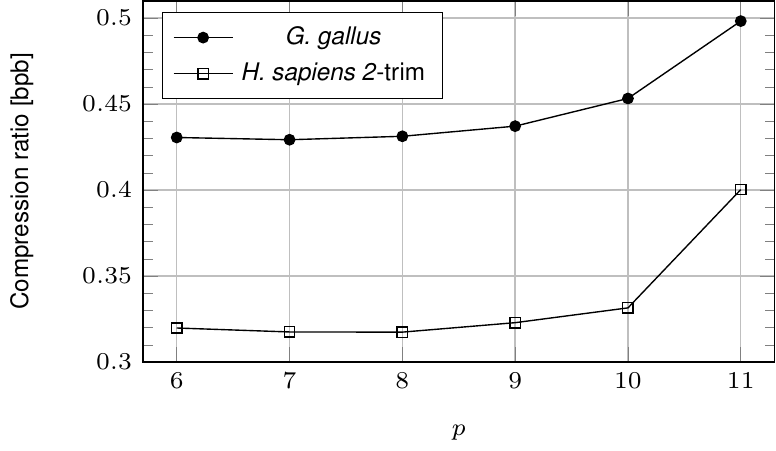}
\else
\centering
\pgfplotsset{width=8.0cm, height=5.25cm}
{\sffamily
\begin{tikzpicture}
\begin{axis}[
	xlabel={$p$},
	ylabel={Compression ratio [bpb]},
	minor y tick num=4,
	minor x tick num=0,
	xmin=5.7,
	ymin=0.3,
	ymax=0.51,
	xmax=11.3,
	grid=major,
	mark size=1.5,
	font=\scriptsize,
	legend entries={\emph{G.\ gallus}, \emph{H.\ sapiens 2}-trim},
	legend columns=1,
	legend pos={north west},
	legend style={font=\scriptsize, mark size=1.5},
]
\addplot[black, mark=*] table[x=sig_len, y=ratio] {dat/sig_len_gg.dat};
\addplot[black, mark=square] table[x=sig_len, y=ratio] {dat/sig_len_hs2.dat};
\end{axis}
\end{tikzpicture}
}\fi
\caption{Compression ratio for various signature lengths for \emph{G.\ gallus} and \emph{H.\ sapiens} 2-trim datasets ($z = 10$)
}
\label{fig:res:sig_k}
\end{figure}
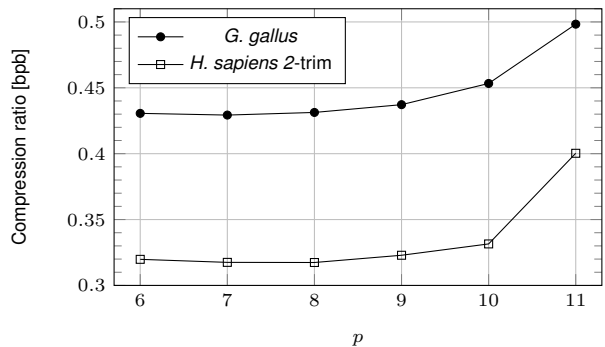

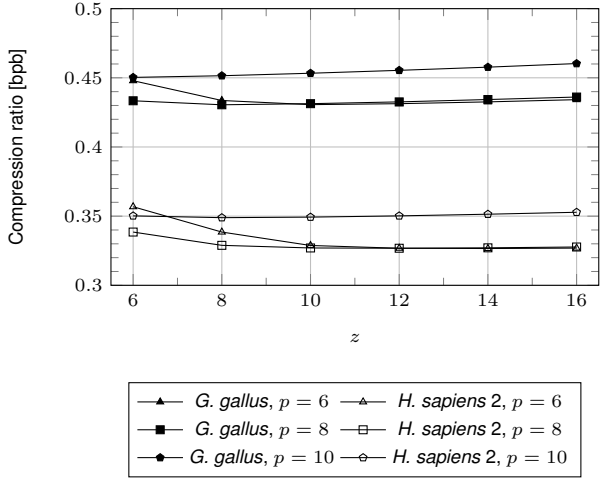
\begin{figure}[t]
\iffigsinpdf
\includegraphics{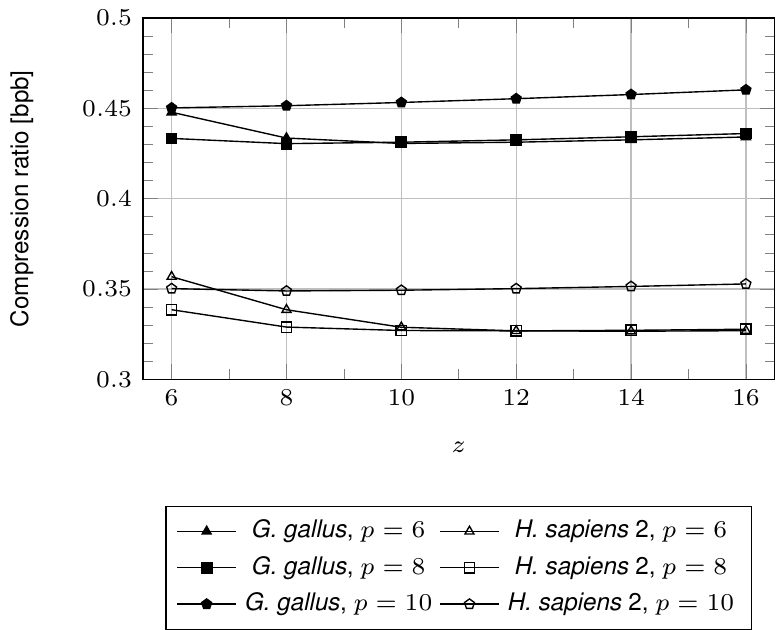}
\else
\centering
\pgfplotsset{width=8.0cm, height=5.25cm}
{\sffamily
\begin{tikzpicture}
\begin{axis}[
	xlabel={$z$},
	ylabel={Compression ratio [bpb]},
	minor y tick num=4,
	minor x tick num=1,
	xmin=5.5,
	ymin=0.3,
	ymax=0.5,
	xmax=16.5,
	grid=major,
	mark size=1.5,
	font=\scriptsize,
	legend entries={
	\emph{G.\ gallus}{,} $p = 6$, 
	\emph{H.\ sapiens} 2{,} $p = 6$, 
	\emph{G.\ gallus}{,} $p = 8$, 
	\emph{H.\ sapiens} 2{,} $p = 8$, 
	\emph{G.\ gallus}{,} $p = 10$, 
	\emph{H.\ sapiens} 2{,} $p = 10$},
	legend style={font=\scriptsize, mark size=1.5, legend columns=2, anchor=north, at={(0.5, -0.35)}},
]
\addplot[black, mark=triangle*] table[x=end_zone, y=GG_6] {dat/end_zone_gg.dat};
\addplot[black, mark=triangle] table[x=end_zone, y=HS2_6] {dat/end_zone_hs2.dat};
\addplot[black, mark=square*] table[x=end_zone, y=GG_8] {dat/end_zone_gg.dat};
\addplot[black, mark=square] table[x=end_zone, y=HS2_8] {dat/end_zone_hs2.dat};
\addplot[black, mark=pentagon*] table[x=end_zone, y=GG_10] {dat/end_zone_gg.dat};
\addplot[black, mark=pentagon] table[x=end_zone, y=HS2_10] {dat/end_zone_hs2.dat};
\end{axis}
\end{tikzpicture}
}\fi
\caption{Compression ratio for various skip zone lengths for \emph{G.\ gallus} and \emph{H.\ sapiens} 2 datasets}
\label{fig:res:zone_k}
\end{figure}

The experimental results with real read data are presented in the upper parts of 
Tables~\ref{tab:results} and~\ref{tab:results2}.
Apart from the proposed compressor ORCOM, we tested 
DSRC~2~\citep{RD2014}, Quip~\citep{JRPK2012}, FQZComp~\citep{BM2013}, 
Scalce~\citep{HNAS2012}, SRcomp~\citep{SC2013} and BWT-SAP~\citep{CBJR2012}.
All these competitors, with the exception of BWT-SAP and SRcomp, are FASTQ compressors, 
and
 all of them present compression results of the separate streams 
In Table~\ref{tab:results} we also present the result of ReCoil~\citep{Y2011} 
on one dataset, copied from~\citep{CBJR2012}.
ReCoil is too slow to be run on all our data in reasonable time.

As we can see in Table~\ref{tab:results}, ORCOM wins on all datasets, 
in an extreme case (\emph{M.\ balbisiana}) with almost twice better 
compression ratio than the second best compressor, BWT-SAP.
Table~\ref{tab:results2} presents the compression times and RAM consumptions.
Our compressor is also usually the fastest, with rather moderate memory usage 
(up to 14\,GB).
We point out that the first phase, distributing the data into bins, is not very costly, 
e.g., it takes less than 25\% of the total time for the largest dataset, 
\emph{H.\ sapiens} 2. 
In the memory use the most frugal is BWT-SAP (which is another 
disk-based software), spending only 3\,MB for each dataset.
One should remember that compression times are related to the number of used 
threads: Quip, FQZComp, SRcomp and BWT-SAP are sequential (1 thread),
while DSRC~2, Scalce and ORCOM are parallel and use 8 threads here.
Moreover, ORCOM, BWT-SAP and SRcomp compress the DNA stream only, while the 
remaining compressors 
have full FASTQ files on the input (with fake remaining streams in case of simulated 
reads presented in Section~\ref{ssec:simul}), what hampers their 
performance in compression speed and memory use (the compression ratios are however 
given for the DNA stream only).
For these reasons, the results from Table~\ref{tab:results2} shouldn't be taken too 
seriously; they are given mostly to point out promising performance of our software.

ORCOM's compression performance depends on two parameters:
the signature length and the skip zone length.
How varying these parameters affects the compression ratio is 
shown in Fig.~\ref{fig:res:sig_k} and Fig.~\ref{fig:res:zone_k}, respectively, 
on the example of two datasets.
It seems that choosing the signature length from $\{6, 7, 8\}$ 
is almost irrelevant for the compression ratio, but with longer signature 
the ratio starts to deteriorate.
The impact of the skip zone length depends somewhat on the chosen signature length, 
yet from our results we can say that any zone length between 8 and 16 
is almost equally good.

\begin{table}
\processtable{Comparison of compression ratios and memory usage between minimizers and signatures in ORCOM.\label{tab:sig}}%
{
\renewcommand{\tabcolsep}{0.3em}
\begin{tabular}{lcccccc}\toprule
Dataset						&&  \multicolumn{2}{c}{Compression ratio}	&& \multicolumn{2}{c}{RAM}\\
\cline{3-4}\cline{6-7}
								&& minimizers	& signatures	&&  minimizers	& signatures	\\
\midrule
\emph{G.\ gallus}			&& 0.443			& 0.433			&&	\q9.7				& \q9.6	\\	
\emph{M.\ balbisiana}	&&	0.112			& 0.110			&&		\q8.3		& \q9.6	\\	
\emph{H. sapiens} 1		&&	1.007				& 1.005			&&	\q7.1				& \q9.7	\\	
\emph{H. sapiens}	2 		&&	0.338			& 0.327			&&	29.7			& 13.8	\\	
\emph{H. sapiens}	2-trim&&	0.330			& 0.317			&&	28.7			& 12.5	\\	
\midrule
\emph{H.\ sapiens} 3		&&	0.188			& 0.175			&&	 26.5			& 11.7	\\	
\emph{H.\ sapiens} 4		&&	0.565			& 0.562			&&	 26.9			& 13.7	\\	
\bottomrule
\end{tabular}
}{}
\end{table}

Finally, we show how replacing the straight minimizers with signatures affects 
the compression ratio and memory consumption (Table~\ref{tab:sig}).
The compression gain is slight, up to 2.3\% on real data 
(\emph{H.\ sapiens} 2) and up to 6.9\% on 
simulated
data (\emph{H.\ sapiens} 3).
Fortunately, the improvement is greater in memory reduction, sometimes exceeding 
factor 2.
As we can see, in two cases using signatures required more memory than with minimizers, 
but this is for two relatively small datasets.
Using signatures generally leads to more even data distribution across bins.

\subsection{Simulated datasets} \label{ssec:simul}
In the experiments for simulated datasets the reads were obtained by randomly sampling \emph{H.\ sapiens} reference genome (HG~37.3).
The number of non N-symbols in the reference is $\sim$2859\,M.
The \emph{H.\ sapiens}~3 dataset contains 1.25\,G reads of length 100\,bp reads 
(to have the genome coverage like for \emph{H.\ sapiens}~2).
Half of them were obtained directly from the reference and another half were reverse-complemented.

The reads in \emph{H.\ sapiens}~4 dataset were obtained from \emph{H.\ sapiens}~3 dataset by modifying each base with a probability 1\% (the probability is independent from the base position).
It is important to stress that such a simulation of errors is far from what happens in real experiments (e.g., in most sequencers the quality of bases depends on the base position).
Nevertheless, this simple error model allows us to compute the theoretically possible compression ratio and compare the results of existing compressors with the estimated optimum and check what 
improvement is still possible in the reads compression area.

The obtained compression ratios are presented in the bottom part of Table~\ref{tab:results}.
We note that ``standard'' FASTQ compressors achieve compression ratios similar 
to the ones on real reads, which is perhaps no surprise.
BWT-SAP achieves a substantial improvement on \emph{H.\ sapiens}~3, 
as the noise in the real data must have broken many long runs in the 
BWT-related sequence and have hampered the compression.
Yet, even more improvement, close to 2-fold, is observed for ORCOM.
This can be explained by the local search for similar reads in our solution: 
once an error affects a read's signature area, the read is moved to another bin.
On the noisy \emph{H.\ sapiens}~4 dataset all compression ratios are, as expected, 
inferior.
It is also not surprising that the standard FASTQ compressors, unable to eliminate
most of the redundancy of the DNA reads, lose less here than ORCOM and Scalce do.
The compression of clean data (\emph{H.\ sapiens}~3) is also faster (Table~\ref{tab:results2}).

It is interesting to compare the ratios with estimations on how good compression ratio is possible.
In theory, it is possible to perform a \emph{de novo} assembly and to reproduce the genome from the reads.

The reads from \emph{H.\ sapiens} 3 dataset can be reordered according to the position of the read in the assembled genome.
Thus, to encode the dataset it is sufficient to encode the assembled genome and for each read also:
\begin{itemize}
\item the position of the read in the assembled genome,
\item the length (in a case of variable-length reads); for our data it is unnecessary as all reads are of length 100\,bp,
\item the read orientation, i.e., whether the read maps the genome directly or must be reverse-complemented.
\end{itemize}

The assembled genome can be encoded using 2 bits per base (there is no N symbols), so for 2859\,Mb we need 5718\,Mbit.
Then, for each read it is necessary to use 1\,bit for the orientation, i.e., 1250\,Mbit in total.
The read positions are ordered but are not unique, so to encode the positions we need to store 
a multisubset of size 1250M from a set of size 2859M, 
which is equivalent to storing 1250M {\em unique} and ordered integers 
from the range $\langle 0, 1250M+2859M) = \langle 0, 4109M)$.

The number of bits necessary to encode $m$ unique and ordered integers from the range $\langle 0, n)$ is
\begin{eqnarray}
&&\log_2 {n \choose m} = \log_2 n! - (\log_2 (n-m)! + \log_2 m!) \approx\nonumber\\
&& \approx n\log_2 n - n - (n-m)\log_2 (n-m) + (n-m) -\nonumber\\
&& - m\log_2 m + m =\nonumber\\ 
&& = n\log_2 n - (n-m)\log_2 (n-m) - m\log_2 m.\qquad\label{eqn:est}
\end{eqnarray}
For $n=4109\times 10^6$ and $m=1250\times 10^6$ we obtain 3642.12\,Mbit.
Thus in total we need 10610.12\,Mbit, so the compression ratio expressed in bits per base is 0.085.

In case of reads with 1\% of wrong bases we need to encode also the bases in the reads and the positions of the differences between the reads and the assembled genome.
There are 1250\,Mbases to encode, but this time it is enough to use $\log_2 3$ bits per base as we are sure that the actual base differs to the base in the genome, so the total size of these data is 1981.20\,Mbit.
To encode the positions, we can conceptually concatenate the reads and encode the ordered and unique positions of wrong bases.
We now apply Eqn.~\ref{eqn:est} with parameters $n=125\times 10^9$ and $m=0.01n$ and obtain 10099.14\,Mbit.
Thus in total, to encode 1.25\,G reads of length 100\,bp with 1\% of wrong bases, we need 22690.46\,Mbit, which translates into 0.182\,bpb.

We can notice that the obtained results with simulated reads are much worse (roughly, 
by factor~2 for \emph{H.\ sapiens}~3 and factor~3 for \emph{H.\ sapiens}~4) than the estimated 
lower bounds.
This is basically due to two reasons. 
One is that the proposed read grouping method belongs to crisp ones, 
i.e., one read belongs to one and only one bin.
In this way, reads with relatively small overlaps are likely to be scattered to different 
bins and their cross-correlation cannot be exploited. 
Moreover, even reads with a large overlap has some (albeit rather small) chance of landing 
in different bins.
This harmful effect of separating similar reads is stronger for noisy data.
The other reason is the simplicity of our modelling, in which read alignment is performed 
only in pairs of reads and thus some long matches may be prematurily truncated.
Overcoming these limitations of our algorithm is an interesting topic for further 
research.

\end{methods}
\section{Conclusions and future work}
We presented ORCOM, a lightweight solution for grouping and 
compressing overlapping reads in DNA sequencing data. 
We showed that the obtained compression ratio 
for large datasets is much better the one from the previously most successful, 
BWT-based, approach.
Our algorithm is based on the recently popular idea of minimizers.
For the human dataset comprising reads of 100\,bp, with 
44.5-fold coverage, we obtain 0.317\,bpb compression ratio. 
This means that the 134.0\,Gb dataset can be stored in as little as 5.31\,GB.
Also for the other tested datasets ORCOM attains, to our knowledge, 
compression ratios better than all previously reported.


ORCOM, as a tool, may be improved in a number of ways. 
Its performance depends, albeit mildly, on two parameters: 
signature length and skip zone length.
Currently their default values are set ad hoc, while in the future 
we are going to work out quite a robust automated parameter selection procedure.
Also, our plans include fine-tuning the backend modeling 
(e.g., the distance function between reads is rather crude now).

More importantly, perhaps, a memory-only mode can be added, 
convenient for powerful machines, 
but with more compact internal data representations affordable also 
for standard PCs, at least on small to moderate sized genomes.
On the other hand, in the disk based mode the memory use may be reduced, 
at least as a trade-off (less RAM, but also fewer threads, thus slower 
compression, and~/~or somewhat worse compression ratio).
From the algorithmic point, a more interesting challenge would be to come 
closer to the compression bounds estimated in Section~\ref{ssec:simul}.
Finally, our ideas could be incorporated in a full-fledged FASTQ compressor, 
together with recent advances in lossy compression of the quality data, 
to obtain unprecedented compression ratios for FASTQ inputs 
in an industry-oriented massively parallel implementation.

\section*{Acknowledgement}
The work was partially supported by the Polish National Science Centre 
under the project DEC-2012/05/B/ST6/03148
and also partially supported by the European Union
from the European Social Fund within the INTERKADRA project
UDAPOKL-04.01.01-00-014/10-00.
The work was performed using the infrastructure supported
by POIG.02.03.01-24-099/13 grant: `GeCONiI—Upper Silesian
Center for Computational Science and Engineering'.

\bibliographystyle{natbib}
\bibliography{orcom}

\newpage
\onecolumn
\setcounter{section}{0}

\centerline{\bfseries\Large Supplementary material for the paper}
\bigskip
\centerline{\bfseries\LARGE\itshape Disk-based genome sequencing data compression}
\bigskip
\centerline{\bfseries\Large by}
\bigskip
\centerline{\bfseries\Large Szymon Grabowski, Sebastian Deorowicz, {\L}ukasz Roguski}
\bigskip
\bigskip
\bigskip
\bigskip


\clearpage
\section{ORCOM usage}
\label{sec:usage}

ORCOM toolkit consists of 2 separate binaries: \textsf{orcom\_bin} for binning DNA records and \textsf{orcom\_pack} for compressing of those bins.

%
%

\subsection{\textsf{orcom\_bin}}
\textsf{orcom\_bin} performs DNA records clustering into separate bins representing signatures. As an input it takes a single or a set of FASTQ files and stores the output to two separate files with binned records: \textsf{*.bdna} file, containing encoded DNA stream, and \textsf{*.bmeta} file, containing archive meta-information.
\bigskip

The general syntax is:
\textsf{orcom\_bin $<$mode$>$ [options]}

where the available \textsf{mode} options are:

\begin{itemize}
\item \textsf{e} --- encode mode,
\item \textsf{d} --- decode mode.
\end{itemize}

\bigskip

In encoding mode, the configuration options are:
\begin{itemize}
\item \textsf{-i$<$file\_name$>$} --- a single input FASTQ file name,
\item \textsf{-o$<$file\_name\_prefix$>$} --- output files name prefix,
\item \textsf{-f"$<$f1$>$ $<$f2$>$ \ldots $<$fn$>$"} --- multiple input FASTQ file list,
\item \textsf{-g} --- input is compressed in gzip format,

\item \textsf{-p$<$value$>$} --- signature length; default: 8,
\item \textsf{-s$<$value$>$} --- record skip zone; default: 12,
\item \textsf{-b$<$value$>$} --- FASTQ input block size (in MB); default: 256,

\item \textsf{-t$<$value$>$} --- number of processing threads; default: no. of available cores.
\end{itemize}

\bigskip

In decoding mode, the configuration options are:
\begin{itemize}
\item \textsf{-i$<$file\_name\_prefix$>$} --- a generated archive input files prefix,
\item \textsf{-o$<$dna\_file$>$} --- output DNA file.

\item \textsf{-t$<$value$>$} --- number of processing threads; default: no. of available cores.
\end{itemize}

\bigskip

The parameters \textsf{-p$<$value$>$} and \textsf{-s$<$value$>$} concern the records clusterization process and signature selection. The parameter \textsf{-b$<$value$>$} concern the bins sizes before and after clusterization---the FASTQ buffer size should be set as large as possible in order to achieve best ratio (at the cost of large memory consumption).
The parameter \textsf{-t$<$value$>$} sets total number of processing threads (not including two I/O threads).

\bigskip 

Here are some usage examples:

\textsf{orcom\_bin e -iNA19238.fastq -oNA19238.bin -t4 -b256 -p6 -s6}

\textsf{orcom\_bin e -f"NA19238\_1.fastq NA19238\_2.fastq" -oNA19238.bin}

\textsf{orcom\_bin e -f"\$( ls *.gz )" -oNA19238.bin -g -t8 -p10 -s14}

\textsf{orcom\_bin d -iNA19238.bin -oNA19238.dna}

%
%

\subsection{\textsf{orcom\_pack}}

\textsf{orcom\_pack} performs DNA records compression.
As an input it takes files produced by \textsf{orcom\_bin}: \textsf{*.bdna} and \textsf{*.bmeta} and it generates two output files: \textsf{*.cdna} file, containing compressed streams and \textsf{*.cmeta} file, containing archive meta-information.

\bigskip

The general syntax is:\\
\textsf{orcom\_pack $<$mode$>$ [options]}

\bigskip

The available \textsf{mode} options are:

\begin{itemize}
\item \textsf{e} --- encode mode,
\item \textsf{d} --- decode mode.
\end{itemize}

\bigskip

In encoding mode the configuration options are:

\begin{itemize}
\item \textsf{-i$<$bin\_files\_prefix$>$} --- input orcom\_bin archive files name prefix,
\item \textsf{-o$<$pack\_files\_prefix$>$} --- output files name prefix,

\item \textsf{-e$<$value$>$} --- encode threshold value; default: 0 (0 - auto),
\item \textsf{-m$<$value$>$} --- mismatch cost: 2,
\item \textsf{-s$<$value$>$} --- insert cost; default: 1,

\item \textsf{-t$<$value$>$} --- number of processing threads; default: no. of available cores.
\end{itemize}

\bigskip

In decoding mode the configuration options are:

\begin{itemize}
\item \textsf{-i$<$pack\_files\_prefix$>$} --- input orcom\_bin archive files name prefix,
\item \textsf{-o$<$dna\_file$>$} --- output DNA file,

\item \textsf{-t$<$value$>$} --- number of processing threads; default: no. of available cores.
\end{itemize}

\bigskip

The parameters \textsf{-e$<$value$>$}, \textsf{-m$<$value$>$} and \textsf{-s$<$value$>$} concern the records internal encoding step, where encoding threshold value should be adapted to the dataset records' length.
The parameter \textsf{-t$<$value$>$} sets total number of processing threads (not including two I/O threads).

\bigskip

Here are some usage examples:

\textsf{orcom\_pack e -iNA19238.bin -oNA19238.orcom -t4}

\textsf{orcom\_pack e -iNA19238.bin -oNA19238.orcom -s2 -m1 -e40}

\textsf{orcom\_pack d -iNA19238.orcom -oNA19238.dna}

\clearpage
\section{Datasets}
\label{sec:datasets}

\subsection{Datasets}
\subsubsection{\emph{G.\ gallus}}
The files were downloaded from the following URLs:\\
\url{ftp://ftp.ddbj.nig.ac.jp/ddbj_database/dra/fastq/SRA030/SRA030308/SRX043656/SRR105788_1.fastq.bz2}\\
\url{ftp://ftp.ddbj.nig.ac.jp/ddbj_database/dra/fastq/SRA030/SRA030308/SRX043656/SRR105788_2.fastq.bz2}\\
\url{ftp://ftp.ddbj.nig.ac.jp/ddbj_database/dra/fastq/SRA030/SRA030309/SRX043656/SRR105789_1.fastq.bz2}\\
\url{ftp://ftp.ddbj.nig.ac.jp/ddbj_database/dra/fastq/SRA030/SRA030309/SRX043656/SRR105789_2.fastq.bz2}\\
\url{ftp://ftp.ddbj.nig.ac.jp/ddbj_database/dra/fastq/SRA030/SRA030312/SRX043656/SRR105792_1.fastq.bz2}\\
\url{ftp://ftp.ddbj.nig.ac.jp/ddbj_database/dra/fastq/SRA030/SRA030312/SRX043656/SRR105792_2.fastq.bz2}\\
\url{ftp://ftp.ddbj.nig.ac.jp/ddbj_database/dra/fastq/SRA030/SRA030314/SRX043656/SRR105794.fastq.bz2}\\
\url{ftp://ftp.ddbj.nig.ac.jp/ddbj_database/dra/fastq/SRA030/SRA030314/SRX043656/SRR105794_1.fastq.bz2}\\
\url{ftp://ftp.ddbj.nig.ac.jp/ddbj_database/dra/fastq/SRA030/SRA030314/SRX043656/SRR105794_2.fastq.bz2}\\
\url{ftp://ftp.ddbj.nig.ac.jp/ddbj_database/dra/fastq/SRA036/SRA036382/SRX043656/SRR197985.fastq.bz2}\\
\url{ftp://ftp.ddbj.nig.ac.jp/ddbj_database/dra/fastq/SRA036/SRA036382/SRX043656/SRR197985_1.fastq.bz2}\\
\url{ftp://ftp.ddbj.nig.ac.jp/ddbj_database/dra/fastq/SRA036/SRA036382/SRX043656/SRR197985_2.fastq.bz2}\\
\url{ftp://ftp.ddbj.nig.ac.jp/ddbj_database/dra/fastq/SRA036/SRA036383/SRX043656/SRR197986.fastq.bz2}\\
\url{ftp://ftp.ddbj.nig.ac.jp/ddbj_database/dra/fastq/SRA036/SRA036383/SRX043656/SRR197986_1.fastq.bz2}\\
\url{ftp://ftp.ddbj.nig.ac.jp/ddbj_database/dra/fastq/SRA036/SRA036383/SRX043656/SRR197986_2.fastq.bz2}\\

Then they were decompressed to a single \url{GG.fastq} file.

\subsubsection{\emph{M. balbisiana}}
The files were downloaded from the following URLs:\\
\url{ftp://ftp.ddbj.nig.ac.jp/ddbj_database/dra/fastq/SRA098/SRA098922/SRX339427/SRR956987.fastq.bz2}\\
\url{ftp://ftp.ddbj.nig.ac.jp/ddbj_database/dra/fastq/SRA098/SRA098922/SRX339427/SRR957627.fastq.bz2}\\

Then they were decompressed to a single \url{MB.fastq} file.

\subsubsection{\emph{H.\ sapiens} 1}
The files were downloaded from the following URLs:\\
\url{ftp://ftp.sra.ebi.ac.uk/vol1/fastq/SRR005/SRR005720/SRR005720_1.fastq.gz}\\
\url{ftp://ftp.sra.ebi.ac.uk/vol1/fastq/SRR005/SRR005720/SRR005720_2.fastq.gz}\\
\url{ftp://ftp.sra.ebi.ac.uk/vol1/fastq/SRR005/SRR005721/SRR005721_1.fastq.gz}\\
\url{ftp://ftp.sra.ebi.ac.uk/vol1/fastq/SRR005/SRR005721/SRR005721_2.fastq.gz}\\
\url{ftp://ftp.sra.ebi.ac.uk/vol1/fastq/SRR005/SRR005734/SRR005734_1.fastq.gz}\\
\url{ftp://ftp.sra.ebi.ac.uk/vol1/fastq/SRR005/SRR005734/SRR005734_2.fastq.gz}\\
\url{ftp://ftp.sra.ebi.ac.uk/vol1/fastq/SRR005/SRR005735/SRR005735_1.fastq.gz}\\
\url{ftp://ftp.sra.ebi.ac.uk/vol1/fastq/SRR005/SRR005735/SRR005735_2.fastq.gz}\\

Then they were decompressed to a single \url{HS1.fastq} file.

\subsubsection{\emph{H.\ sapiens} 2}
The FASTQ files (48 files) were downloaded from the following URLs:\\
\url{ftp://ftp.sra.ebi.ac.uk/vol1/fastq/ERR024/ERR024163/ERR024163_1.fastq.gz}\\
\url{ftp://ftp.sra.ebi.ac.uk/vol1/fastq/ERR024/ERR024163/ERR024163_2.fastq.gz}\\
\url{ftp://ftp.sra.ebi.ac.uk/vol1/fastq/ERR024/ERR024164/ERR024164_1.fastq.gz}\\
\url{ftp://ftp.sra.ebi.ac.uk/vol1/fastq/ERR024/ERR024164/ERR024164_2.fastq.gz}\\
\url{ftp://ftp.sra.ebi.ac.uk/vol1/fastq/ERR024/ERR024165/ERR024165_1.fastq.gz}\\
\url{ftp://ftp.sra.ebi.ac.uk/vol1/fastq/ERR024/ERR024165/ERR024165_2.fastq.gz}\\
\url{ftp://ftp.sra.ebi.ac.uk/vol1/fastq/ERR024/ERR024166/ERR024166_1.fastq.gz}\\
\url{ftp://ftp.sra.ebi.ac.uk/vol1/fastq/ERR024/ERR024166/ERR024166_2.fastq.gz}\\
\url{ftp://ftp.sra.ebi.ac.uk/vol1/fastq/ERR024/ERR024167/ERR024167_1.fastq.gz}\\
\url{ftp://ftp.sra.ebi.ac.uk/vol1/fastq/ERR024/ERR024167/ERR024167_2.fastq.gz}\\
\url{ftp://ftp.sra.ebi.ac.uk/vol1/fastq/ERR024/ERR024168/ERR024168_1.fastq.gz}\\
\url{ftp://ftp.sra.ebi.ac.uk/vol1/fastq/ERR024/ERR024168/ERR024168_2.fastq.gz}\\
\url{ftp://ftp.sra.ebi.ac.uk/vol1/fastq/ERR024/ERR024169/ERR024169_1.fastq.gz}\\
\url{ftp://ftp.sra.ebi.ac.uk/vol1/fastq/ERR024/ERR024169/ERR024169_2.fastq.gz}\\
\url{ftp://ftp.sra.ebi.ac.uk/vol1/fastq/ERR024/ERR024170/ERR024170_1.fastq.gz}\\
\url{ftp://ftp.sra.ebi.ac.uk/vol1/fastq/ERR024/ERR024170/ERR024170_2.fastq.gz}\\
\url{ftp://ftp.sra.ebi.ac.uk/vol1/fastq/ERR024/ERR024171/ERR024171_1.fastq.gz}\\
\url{ftp://ftp.sra.ebi.ac.uk/vol1/fastq/ERR024/ERR024171/ERR024171_2.fastq.gz}\\
\url{ftp://ftp.sra.ebi.ac.uk/vol1/fastq/ERR024/ERR024172/ERR024172_1.fastq.gz}\\
\url{ftp://ftp.sra.ebi.ac.uk/vol1/fastq/ERR024/ERR024172/ERR024172_2.fastq.gz}\\
\url{ftp://ftp.sra.ebi.ac.uk/vol1/fastq/ERR024/ERR024173/ERR024173_1.fastq.gz}\\
\url{ftp://ftp.sra.ebi.ac.uk/vol1/fastq/ERR024/ERR024173/ERR024173_2.fastq.gz}\\
\url{ftp://ftp.sra.ebi.ac.uk/vol1/fastq/ERR024/ERR024174/ERR024174_1.fastq.gz}\\
\url{ftp://ftp.sra.ebi.ac.uk/vol1/fastq/ERR024/ERR024174/ERR024174_2.fastq.gz}\\
\url{ftp://ftp.sra.ebi.ac.uk/vol1/fastq/ERR024/ERR024175/ERR024175_1.fastq.gz}\\
\url{ftp://ftp.sra.ebi.ac.uk/vol1/fastq/ERR024/ERR024175/ERR024175_2.fastq.gz}\\
\url{ftp://ftp.sra.ebi.ac.uk/vol1/fastq/ERR024/ERR024176/ERR024176_1.fastq.gz}\\
\url{ftp://ftp.sra.ebi.ac.uk/vol1/fastq/ERR024/ERR024176/ERR024176_2.fastq.gz}\\
\url{ftp://ftp.sra.ebi.ac.uk/vol1/fastq/ERR024/ERR024177/ERR024177_1.fastq.gz}\\
\url{ftp://ftp.sra.ebi.ac.uk/vol1/fastq/ERR024/ERR024177/ERR024177_2.fastq.gz}\\
\url{ftp://ftp.sra.ebi.ac.uk/vol1/fastq/ERR024/ERR024178/ERR024178_1.fastq.gz}\\
\url{ftp://ftp.sra.ebi.ac.uk/vol1/fastq/ERR024/ERR024178/ERR024178_2.fastq.gz}\\
\url{ftp://ftp.sra.ebi.ac.uk/vol1/fastq/ERR024/ERR024179/ERR024179_1.fastq.gz}\\
\url{ftp://ftp.sra.ebi.ac.uk/vol1/fastq/ERR024/ERR024179/ERR024179_2.fastq.gz}\\
\url{ftp://ftp.sra.ebi.ac.uk/vol1/fastq/ERR024/ERR024180/ERR024180_1.fastq.gz}\\
\url{ftp://ftp.sra.ebi.ac.uk/vol1/fastq/ERR024/ERR024180/ERR024180_2.fastq.gz}\\
\url{ftp://ftp.sra.ebi.ac.uk/vol1/fastq/ERR024/ERR024181/ERR024181_1.fastq.gz}\\
\url{ftp://ftp.sra.ebi.ac.uk/vol1/fastq/ERR024/ERR024181/ERR024181_2.fastq.gz}\\
\url{ftp://ftp.sra.ebi.ac.uk/vol1/fastq/ERR024/ERR024182/ERR024182_1.fastq.gz}\\
\url{ftp://ftp.sra.ebi.ac.uk/vol1/fastq/ERR024/ERR024182/ERR024182_2.fastq.gz}\\
\url{ftp://ftp.sra.ebi.ac.uk/vol1/fastq/ERR024/ERR024183/ERR024183_1.fastq.gz}\\
\url{ftp://ftp.sra.ebi.ac.uk/vol1/fastq/ERR024/ERR024183/ERR024183_2.fastq.gz}\\
\url{ftp://ftp.sra.ebi.ac.uk/vol1/fastq/ERR024/ERR024184/ERR024184_1.fastq.gz}\\
\url{ftp://ftp.sra.ebi.ac.uk/vol1/fastq/ERR024/ERR024184/ERR024184_2.fastq.gz}\\
\url{ftp://ftp.sra.ebi.ac.uk/vol1/fastq/ERR024/ERR024185/ERR024185_1.fastq.gz}\\
\url{ftp://ftp.sra.ebi.ac.uk/vol1/fastq/ERR024/ERR024185/ERR024185_2.fastq.gz}\\
\url{ftp://ftp.sra.ebi.ac.uk/vol1/fastq/ERR024/ERR024186/ERR024186_1.fastq.gz}\\
\url{ftp://ftp.sra.ebi.ac.uk/vol1/fastq/ERR024/ERR024186/ERR024186_2.fastq.gz}\\

Then they were decompressed to a single \url{HS2.fastq} file.

\subsubsection{\emph{H.\ sapiens} 2-trim}
The FASTQ file was obtained from \emph{H.\ sapiens} 2 dataset by truncating all reads to 100\,bp.

\subsubsection{\emph{H.\ sapiens} 3}
The FASTQ file was obtained by sampling the reference genome without introducing any errors. The FASTA reference file was downloaded from 1000 Genomes Project (\url{ftp://ftp.1000genomes.ebi.ac.uk/vol1/ftp/technical/reference/human_g1k_v37.fasta.gz}) and the final reference \emph{std\_ref.fasta} was built using all 22 standard chromosomes, X~and Y sex chromosomes, and mitochondrial MT sequence.

\bigskip

The FASTQ file for this experiment was generated using \emph{gen\_fastq} tool with 1250 million of reads and \nobreak{100 bp} length as parameters:

\textsf{./gen\_fastq 1250000000 100 std\_ref.fasta HS3.fastq}

\subsubsection{\emph{H.\ sapiens} 4}
The FASTQ file was obtained from \emph{H.\ sapiens} 3 by modifying each base with an error probability of 1\%.

\clearpage

\section{Tests configurations}
\label{sec:test}

\subsection{Test platform}
ORCOM toolset was implemented in C++ and compiled using g++ compiler (version 4.8.2) for the linux build.

The configuration of the test machine was:
\begin{itemize}
\item CPU: AMD Opteron 6136 (8-cores clocked at 2.4\,GHz),
\item RAM: 128\,GB RAM,
\item HDD: RAID-5 disk matrix containing 6 HDDs. 
\end{itemize}

\subsection{Applications}

\subsubsection*{DSRC~2}
DSRC 2 (ver.\ 2.00) was executed with 8 threads and DNA compression level 2.

\bigskip

Command line:\\
\textsf{dsrc c -m2 -t8 $<$input\_fastq\_file$>$ $<$output\_dsrc\_file$>$}

\subsubsection*{Quip}
Quip (ver.\ 1.1.7) was executed in assembly and verbose mode (to collect stream sizes information).

\bigskip

Command line:\\
\textsf{quip -a -v $<$input\_fastq\_file$>$}

\subsubsection*{FQZComp}
FQZComp (ver.\ 4.6) was executed with maximum sequence compression level and using both strands for compression table.

\bigskip

Command line:\\
\textsf{fqz\_comp -s8+ -b $<$input\_fastq\_file$>$ $<$output\_fqzcomp\_file$>$}

\subsubsection*{Scalce}
Scalce (ver.\ 2.7) was executed with 8 threads. 

\bigskip

Command line:\\
\textsf{scalce $<$input\_fastq\_file$>$ -T 8 -o $<$output\_scalce\_files\_prefix$>$}

\subsubsection*{SRComp}
Before running SRComp the DNA stream was extracted from FASTQ files, as it's the only input format \nobreak{SRComp} accepts.

\bigskip

Command line:\\
\textsf{SRComp -e $<$input\_dna\_file$>$}

\subsubsection*{BWT-SAP}
BEETL framework was used in 0.10 version, where \textsf{bwt} subtool was used in BWT-SAP mode with external BCR construction algorithm (as the in-memory mode was not yet working properly).

\bigskip

Command line:\\
\textsf{beetl bwt --output-format=ASCII --sap-ordering --algorithm=ext -i $<$input\_fastq\_file$>$ -o $<$output\_bwt\_files\_prefix$>$}

\subsubsection*{ORCOM}
ORCOM toolset was used in 1.0a version with 8 processing threads, 256 MB of input FASTQ block size (default) and parameters depending on the used dataset.

\begin{center}
    \begin{tabular}{lcccccc}\toprule
    Dataset 		& Signature length  	& Skip-zone length 	 	& Encode threshold \\ \midrule
    G. gallus 		& 8 				    & 12 					& 50 \\ 
    M. balbisiana 	& 8 					& 12 					& 50 \\ 
    H. sapiens 1 	& 6 					& 6 					& 20 \\ 
    H. sapiens 2 	& 8                     & 12 					& 50 \\ 
    H. sapiens 2 - trim	& 8 			    & 12 					& 50 \\ \midrule
    H. sapiens 3	& 8 	   				& 12 					& 50 \\ 
    H. sapiens 4	& 8 	 				& 12 					& 50 \\ 
\bottomrule
    \end{tabular}
\end{center}

\bigskip

Command lines (main tests):\\
\textsf{orcom\_bin e -t8 -p$<$signature\_length$>$ -s$<$skipzone\_length$>$ -i$<$input\_fastq\_file$>$ -o$<$output\_bin\_file$>$}\\
\textsf{orcom\_pack e -t8 -e$<$encode\_threshold$>$ -i$<$input\_bin\_file$>$ -o$<$output\_orcom\_file$>$}

\end{document}